\documentclass[superscriptaddress,preprint,amsmath,amssymb,aps,prl]{revtex4-2}

\usepackage{graphicx}
\usepackage{amsmath}
\usepackage{dcolumn}
\usepackage{txfonts}
\usepackage{bm}
\usepackage{mathrsfs}
\usepackage{amsfonts}
\usepackage{amssymb}
\usepackage{braket}
\usepackage{color}
\usepackage[colorlinks=true,citecolor=magenta,anchorcolor=blue]{hyperref}

\begin{document}

\title{Higher-order topological in-bulk corner state in pure diffusion systems}

\author{Zhoufei Liu}
\affiliation{Department of Physics, State Key Laboratory of Surface Physics, and Key Laboratory of Micro and Nano Photonic Structures (MOE), Fudan University, Shanghai 200438, China}

\author{Pei-Chao Cao}
\affiliation{State Key Laboratory of Extreme Photonics and Instrumentation, ZJU-Hangzhou Global Scientific and Technological Innovation Center, Zhejiang University, Hangzhou 310027, China}
\affiliation{International Joint Innovation Center, Key Lab. of Advanced Micro/Nano Electronic Devices $\&$ Smart Systems of Zhejiang, The Electromagnetics Academy of Zhejiang University, Zhejiang University, Haining 314400, China}
\affiliation{Shaoxing Institute of Zhejiang University, Zhejiang University, Shaoxing 312000, China}

\author{Liujun Xu}
\affiliation{Graduate School of China Academy of Engineering Physics, Beijing 100193, China}

\author{Guoqiang Xu}
\affiliation{Department of Electrical and Computer Engineering, National University of Singapore, Kent Ridge 117583, Republic of Singapore}

\author{Ying Li}\email{eleying@zju.edu.cn}
\affiliation{State Key Laboratory of Extreme Photonics and Instrumentation, ZJU-Hangzhou Global Scientific and Technological Innovation Center, Zhejiang University, Hangzhou 310027, China}
\affiliation{International Joint Innovation Center, Key Lab. of Advanced Micro/Nano Electronic Devices $\&$ Smart Systems of Zhejiang, The Electromagnetics Academy of Zhejiang University, Zhejiang University, Haining 314400, China}
\affiliation{Shaoxing Institute of Zhejiang University, Zhejiang University, Shaoxing 312000, China}

\author{Jiping Huang}\email{jphuang@fudan.edu.cn}
\affiliation{Department of Physics, State Key Laboratory of Surface Physics, and Key Laboratory of Micro and Nano Photonic Structures (MOE), Fudan University, Shanghai 200438, China}

\date{\today}

\begin{abstract}

Compared with conventional topological insulator that carries topological state at its boundaries, the higher-order topological insulator exhibits lower-dimensional gapless boundary states at its corners and hinges. Leveraging the form similarity between Schr${\rm{\ddot{o}}}$dinger equation and diffusion equation, researches on higher-order topological insulators have been extended from condensed matter physics to thermal diffusion. Unfortunately, all the corner states of thermal higher-order topological insulator reside within the band gap. Another kind of corner state, which is embedded in the bulk states, has not been realized in pure diffusion systems so far. Here, we construct higher-dimensional Su-Schrieffer-Heeger models based on sphere-rod structure to elucidate these corner states, which we term ``in-bulk corner states". Due to the anti-Hermitian properties of diffusive Hamiltonian, we investigate the thermal behaviour of these corner states through theoretical calculation, simulation, and experiment. Furthermore, we study the different thermal behaviours of in-bulk corner state and in-gap corner state. Our results would open a different gate for diffusive topological states and provide a distinct application for efficient heat dissipation.

\end{abstract}

\maketitle

{\it Introduction.}--Over the last several decades, topological materials have garnered significant interest from the community of condensed matter physics~\cite{HasanRMP10, QiRMP11}. Since the discovery of quantum Hall effect in the 1980s~\cite{KlitzingPRL80}, numerous exotic topological phases have been theoretically predicted and experimentally discovered, such as quantum spin Hall insulators~\cite{KanePRL05}, quantum anomalous Hall insulators~\cite{ChangSci13, CaoPER22}, and Weyl semimetals~\cite{WanPRB11}. Recently, a novel class of topological insulators, known as higher-order topological insulators, has been proposed beyond the conventional bulk-boundary correspondence~\cite{BenalcazarSci17, SchindlerSA18}. For example, an $m$-dimensional ($m$D) topological insulator with ($m-n$)D topological boundary states is called the $n$th-order topological insulator. Except for the real material systems such as bismuth~\cite{SchindlerNP18}, the higher-order topological insulator has been realized in various classical wave systems, such as photonics~\cite{MittalNP19, HeNC20}, acoustics~\cite{XueNM19, NiNM19}, and mechanics~\cite{GarciaNat18}.  
  
Beyond waves, diffusion is another crucial mechanism for energy and mass transfer. For flexible heat regulation, thermal metamaterials~\cite{Huang20, Xu23, YangPR21, LiNRM21, ZhangNRP23, YangRMP24, JuAM23, XuPRE18, ShenAPL16} have been designed to realize novel functions, including thermal cloaking~\cite{FanAPL08, Yeung22, JinPNAS23}, thermal coding~\cite{HuAM19, GuoAM22, JinAM24, JinRes23}, spatiotemporal nonreciprocity~\cite{XuPRL22-1, LiNC22}, and thermal sensor~\cite{JinIJHMT20}. Currently, heat diffusion has been shown as an exotic platform to implement the topological phenomena and non-Hermitian physics~\cite{Liuarxiv23-1}, such as exceptional point encirclement~\cite{XuPRL21}, Weyl exceptional ring~\cite{XuPNAS22}, topological edge state~\cite{XuNP22, YoshidaSR21, QiAM22, HuAM22}, non-Hermitian skin effect~\cite{CaoCP21, CaoCPL22, Liuarxiv23-2}, parity-time symmetry~\cite{Caoarxiv23}, and quasicrystals~\cite{Liuarxiv22}. Besides, higher-order topological insulators have been realized in thermal diffusion recently~\cite{XuNC23, WuAM23, FukuiPRE23, Qiarxiv23, Chenarxiv23}, but all the corner states in these works are localized in the band gap. The challenge remains to realize the corner state embedded in the bulk in thermal diffusion, while it has been realized in classical wave systems as a bound state in the continuum~\cite{CerjanPRL20, HuLSA21, WangLSA21, LiuPRL23, QianPRL24}. 

Here, we design and experimentally demonstrate a higher-dimensional heat conduction model with sphere-rod structure to construct purely diffusive 2D Su–Schrieffer–Heeger (SSH) model~\cite{ZhengPER23, LiPER23}. This diffusive higher-dimensional SSH model is a second-order topological insulator with corner states embedded in the bulk, which are called ``in-bulk corner state" in this work. Due to the dissipation nature of this model, we study the thermal behaviour of these in-bulk corner states in theoretical calculation, simulation, and experiment. Besides, we compare the different thermal behaviours between in-bulk corner state and in-gap corner state. The in-bulk corner state has a fast decay rate and a localized temperature field, which could help intensify the heat dissipation without affecting the neighbourhood.

\begin{figure}[!ht]
\includegraphics[width=\linewidth]{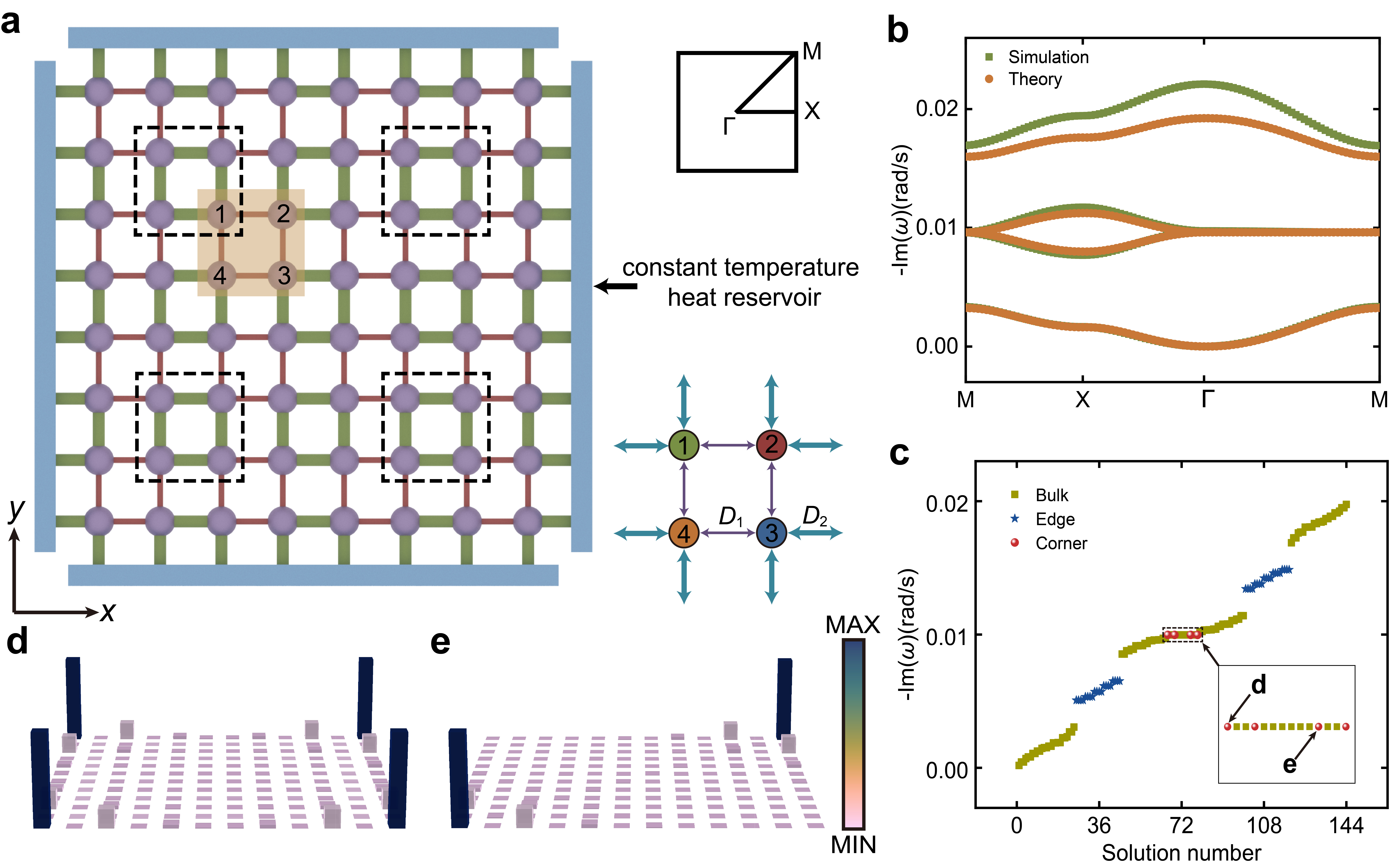}
\caption{Diffusive 2D SSH model with in-bulk corner states. (a) Schematic diagram of sphere-rod structure with four constant temperature boundaries set at the room temperature (i.e., fixed boundary condition). Thermal diffusivities of red thin and green thick rods are $D_{1}$ and $D_{2}$. Blue bars denote constant temperature heat reservoirs. The light brown box indicates one unit cell with four spheres. The upper inset is the first Brillouin zone of diffusive 2D SSH model. The lower inset is schematic diagram for the equivalent tight-binding model of one unit cell. The spheres in the black dashed boxes are simulated as the bulk excitation discussed in Fig.~\ref{Fig4}. (b) The band structure of diffusive 2D SSH model. (c) The fixed boundary condition spectrum of diffusive 2D SSH model. The inset is the enlarged view of four in-bulk corner states and their adjacent bulk states. (d,e) The mode distributions of two in-bulk corner states, which are indicated in the inset of Fig.~\ref{Fig1}(c).} 
\label{Fig1}
\end{figure}

{\it Diffusive 2D SSH model based on the sphere-rod structure.}--In the condensed matter physics, the 2D SSH model is recognized as a higher-order topological insulator with dipole moment~\cite{XiePRB18}. To implement the 2D SSH model in diffusion systems, we construct a sphere-rod structure with the same material parameter and length of rods [see the schematic diagram in Fig.~\ref{Fig1}(a)]. Analogous to different hopping amplitudes in 2D SSH model, we vary the radius of rods to implement different thermal diffusivities. The radii of intracell and intercell rods are denoted as $R_{0,1}$ and $R_{0,2}$. To maintain a constant onsite diffusivity for the boundary spheres, we impose the fixed boundary condition by connecting the boundary rods with constant temperature heat reservoirs. The discretized thermal lattice system yields an effective fixed boundary condition Hamiltonian for the sphere-rod structure, serving as the diffusive counterpart of the 2D SSH model (detailed derivation in Supplementary Note I~\cite{Supp}). After performing the Fourier transformation, the Bloch Hamiltonian can be written as 
\begin{equation}
{\hat{H}}(\bm{k})=\
(-i)\left(\begin{matrix}
        2\left(D_{1}+D_{2}\right) & -D_{1}-D_{2}e^{ik_{x}} & 0 & -D_{1}-D_{2}e^{ik_{y}}
        \\
        -D_{1}-D_{2}e^{-ik_{x}} & 2\left(D_{1}+D_{2}\right) & -D_{1}-D_{2}e^{ik_{y}} & 0
        \\
        0 & -D_{1}-D_{2}e^{-ik_{y}} & 2\left(D_{1}+D_{2}\right) & -D_{1}-D_{2}e^{-ik_{x}}
        \\
        -D_{1}-D_{2}e^{-ik_{y}} & 0 & -D_{1}-D_{2}e^{ik_{x}} & 2\left(D_{1}+D_{2}\right)\\
\end{matrix}\right)
\end{equation}
where $D_{1}$ and $D_{2}$ are thermal diffusivities of intracell and intercell rods. The values of thermal diffusivities are obtained from the fitting between simulated eigenvalues of the sphere-rod structure and theoretical eigenvalues of the tight-binding model [see the band structure in Fig.~\ref{Fig1}(b)]. The band structure reveals two band gaps in the diffusive 2D SSH model: one between the first and second bands, and another between the third and fourth bands. Besides, the second and third bands hybridize with each other, which gives an opportunity for the higher-order in-bulk corner states. Here the eigenvalue $\omega$ in thermal diffusion is purely imaginary, demonstrating the dissipation nature of temperature field. The topological invariant of this higher-order topological insulator is demonstrated in Supplementary Note II~\cite{Supp}.

Figure~\ref{Fig1}(c) shows the fixed boundary condition spectrum of diffusive 2D SSH model. We can find that four thermal corner states are embedded within, rather than hybridizing with, the central bulk states. We call them as ``in-bulk corner states" in this work. Furthermore, the eigenstate distributions of in-bulk corner states are confined at four or two corners of the structure with a notable localization [see Figs.~\ref{Fig1}(d,e)]. 

{\it Experimental observation of in-bulk corner state.}--Next, we perform the experiment to observe these in-bulk corner states in the sphere-rod structure. The schematic diagram of experimental setup is shown in Fig.~\ref{Fig2}(a) [see the image of experimental setup in Supplementary Fig.~S13(a)~\cite{Supp} and photo of fabricated nontrivial sample in Supplementary Fig.~S13(b)~\cite{Supp}]. To facilitate the observation in the experiment, we choose to excite only one corner sphere and apply the fixed boundary condition along two boundaries intersecting at this corner sphere. This configuration has broken the $C_{4v}$ symmetry due to the change of boundary conditions. In this case, only one thermal corner state is embedded in the bulk, which maintains the same decay rate as those preserving the $C_{4v}$ symmetry [see Supplementary Fig.~S2(a)~\cite{Supp}]. Besides, the distribution of this in-bulk corner state is localized at this excited corner sphere [see Supplementary Fig.~S2(b)~\cite{Supp}]. The other three thermal corner states will shift to hybridize with the thermal edge states. We capture the temperature evolution of this corner sphere to investigate the thermal behaviour of in-bulk corner state. As a reference for the experiment, we theoretically solve a series of partial differential equations (Supplementary Note I~\cite{Supp}) and perform the temperature field simulation by COMSOL Multiphysics (Supplementary Note XI~\cite{Supp}). The theoretical and simulated results show a good agreement with the experimental data. In the nontrivial phase, the temperature evolution of corner sphere exhibits an exponential decay predicted by the decay rate of corner state [${\rm{-Im}}(\omega)_{\rm{corner}}=0.0100~{\rm{rad/s}}$, see Fig.~\ref{Fig2}(b)]. Besides, the temperature field for corner sphere introduces the smallest heating to the adjacent spheres with a localized distribution [see Figs.~\ref{Fig2}(c-h)]. 

In contrast, for the trivial lattice without corner state, the temperature evolution of corner sphere deviates from the nontrivial results, failing to exhibit an exponential decay [see Fig.~\ref{Fig2}(b)]. The trivial lattice displays an extended temperature field distribution [see Figs.~\ref{Fig3}(a-f)]. As a comparison, the nontrivial temperature field decays faster and has a more confined distribution, advantageous for the efficient heat dissipation without affecting the surrounding. Additionally, we investigate the thermal behaviour of bulk state (Supplementary Note IV~\cite{Supp}) and edge state (Supplementary Note V~\cite{Supp}). The thermal edge state also shows a localized temperature field. 

These in-bulk corner states are sensitive to alterations in single geometric or material parameter (Supplementary Note VI~\cite{Supp}). However, these states are robust when simultaneously adjusting two parameters in proportion (Supplementary Note VI~\cite{Supp}). Besides, these in-bulk corner states have a robustness against different numbers of unit cells (Supplementary Note VI~\cite{Supp}). The decay rate of in-bulk corner state stands still under different environmental temperatures but increases with a stronger environmental convection (Supplementary Note VII~\cite{Supp}). Moreover, these in-bulk corner states are robust against defects with a fixed decay rate (Supplementary Note VIII~\cite{Supp}). 

{\it Comparison between in-bulk corner state and in-gap corner state.}--As depicted in Fig.~\ref{Fig2}(a), the diffusive 2D SSH model previously discussed is isotropic with $R_{0,1x}(R_{0,2x})=R_{0,1y}(R_{0,2y})$. However, when the radii are different along two directions ($R_{0,2x}{\neq}R_{0,2y}$), this model becomes anisotropic with the $C_{2v}$ symmetry rather than the $C_{4v}$ symmetry [see the schematic diagram in Fig.~\ref{Fig4}(a)]. This anisotropic model also acts as a higher-order topological insulator with nonzero polarizations along two directions (Supplementary Note IX~\cite{Supp}). The thermal corner states of anisotropic diffusive 2D SSH model are lying in the band gap, which we call ``in-gap corner states" [see the fixed boundary condition spectrum in Fig.~\ref{Fig4}(b)]. Similarly, a box-shaped combined configuration of diffusive 2D SSH model also exhibits in-gap corner states (Supplementary Note X~\cite{Supp})~\cite{WuAM23, FukuiPRE23, Qiarxiv23, Chenarxiv23}. The different characteristics of in-bulk corner state and in-gap corner state will lead to different thermal behaviours, which are shown as follows. We choose two initial conditions to stimulate the temperature field. The first is to excite four corner spheres, and the other is to excite four corner spheres and some bulk spheres simultaneously [bulk spheres marked in the black dashed box of Figs.~\ref{Fig1}(a) and~\ref{Fig4}(a), Supplementary Note XI~\cite{Supp}]. For the in-bulk corner state, the decay rates of some bulk states are the same as or close to that of corner state. Therefore the temperature evolution of corner sphere under these two excitations are close with each other. However, for the in-gap corner state, there is a significant discrepancy between the decay rates of bulk and corner states. The bulk excitation will have a stronger influence on the temperature field of corner spheres. So the temperature evolution of corner sphere under these two excitations will differ markedly. Indeed, with different $T_{\rm{bulk}}$s, the decay rate difference (decay rate defined here as the rate of corner sphere's normalized temperature ${\rm{log}}[(T-T_{f})/(T_{i}-T_{f})]$ over time $t$, rather than retrieved from the fixed boundary condition spectrum) of in-gap corner state under two excitations is more pronounced than the one of in-bulk corner state, as shown in Fig.~\ref{Fig4}(c).

Furthermore, we investigate the temperature field behaviours of in-bulk corner state and in-gap corner state by adjusting the geometric parameter $R_{0,2y}$. As before, we employ two initial conditions: corner spheres excitation and corner and bulk spheres excitation. Besides, we fix the same $T_{\rm{bulk}}$ at this time. With $R_{0,2y}>R_{0,2x}$, the thermal corner states become the in-gap states rather than the in-bulk states. Besides, the gap between the second and third bands widens as $R_{0,2y}$ increases [see the inset of Fig.~\ref{Fig4}(d)]. As a result, the decay rate of bulk states from the fixed boundary condition spectrum will deviate larger from the corner states. Therefore, the decay rate difference under two excitations will become larger with the increase of $R_{0,2y}$ [see Fig.~\ref{Fig4}(d)]. Above discussion about the temperature field behaviours of in-bulk corner state and in-gap corner state highlights the difference between these two corner states.

{\it Conclusion.}--In summary, we have realized the higher-order topological insulator in pure heat conduction systems by constructing diffusive 2D SSH model. This model exhibits four thermal corner states embedded in the bulk states rather than lying in the band gap. Theoretical calculation, finite-element simulations, and experiment confirm the existence of these in-bulk corner states. The in-bulk corner state shows a rapidly decaying and localized temperature field under the corner sphere excitation. Furthermore, we present a comparison between the distinct thermal behaviours of in-bulk and in-gap corner states. Our results will motivate the exploration of more topological states of matter in pure diffusion systems~\cite{LiuCPL23}. Besides, our findings would bring new insights into robust heat manipulation and provide a distinct mechanism for heat dissipation.

{\it Acknowledgments.}--We express our gratitude to Prof. Ruibao Tao for the insightful discussions. J. H. gratefully acknowledges the support received from the National Natural Science Foundation of China through Grants No. 12035004 and No. 12320101004, as well as the Innovation Program of the Shanghai Municipal Education Commission with Grant No. 2023ZKZD06. Y. L. acknowledges the funding support from the National Natural Science Foundation of China under Grants Nos. 92163123 and 52250191, along with the Zhejiang Provincial Natural Science Foundation of China with Grant No. LZ24A050002. P. C. is grateful for the support provided by the China Postdoctoral Science Foundation through Grant 2023M733120. L. X. acknowledges the support received from the National Natural Science Foundation of China under Grants No. 12375040, No. 12088101, and No. U2330401.

\clearpage
\newpage

\clearpage
\newpage
\begin{figure}[!ht]
\includegraphics[width=\linewidth]{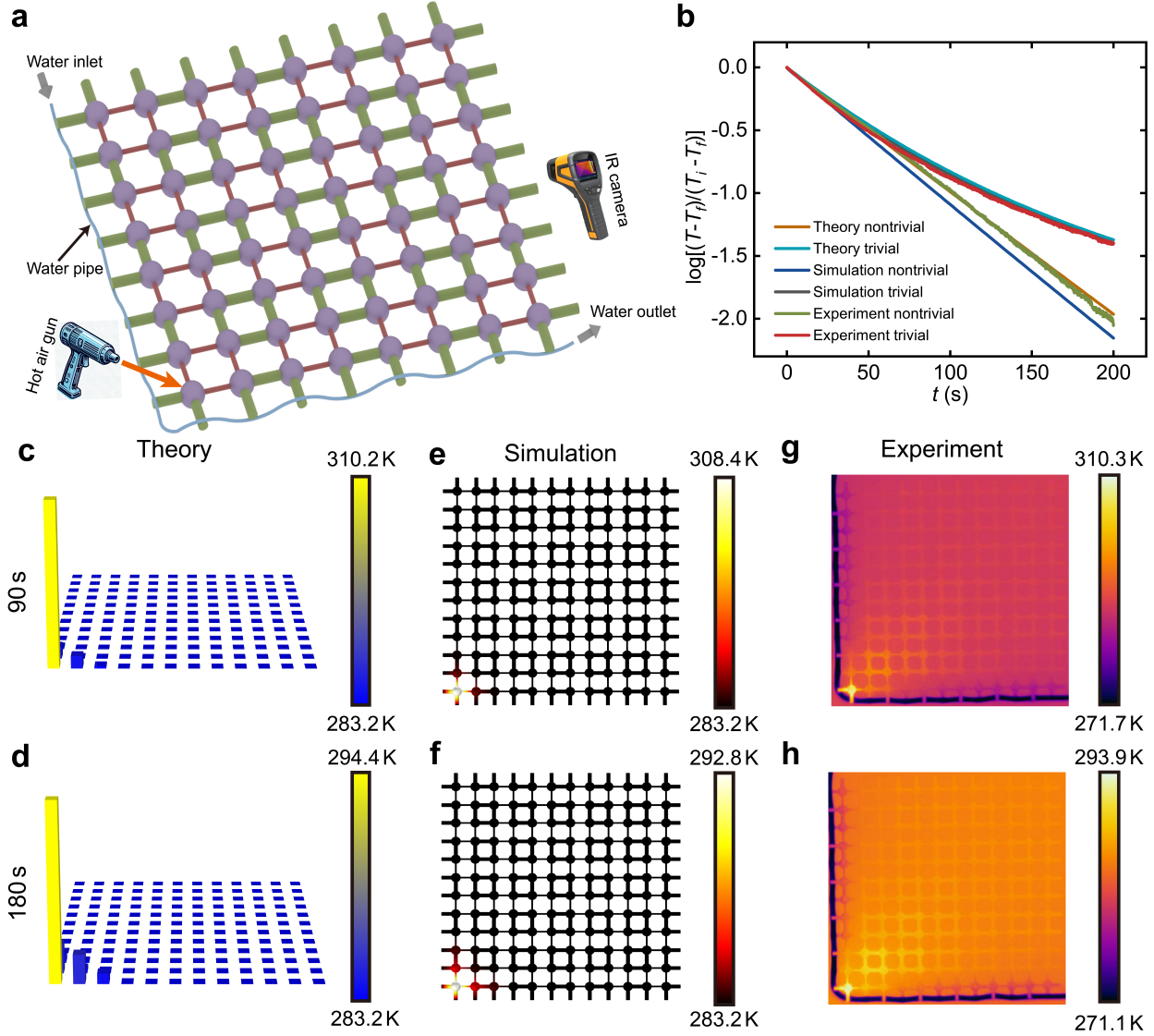}
\caption{In-bulk corner state in the nontrivial lattice. (a) Schematic diagram of the experimental setup. (b) Normalized temperature evolution of corner sphere for the nontrivial and trivial lattices. Here $T_{i}$ is the excited temperature of corner sphere and $T_{f}$ is the temperature at the thermal equilibrium, i.e., room temperature. (c,d) Theoretical temperature distributions for the nontrivial lattice at (c) 90~s and (d) 180~s. (e,f) Simulated temperature distributions for the nontrivial lattice at (e) 90~s and (f) 180~s. (g,h) Experimental temperature distributions for the nontrivial lattice at (g) 90~s and (h) 180~s.}
\label{Fig2}
\end{figure}

\clearpage
\newpage
\begin{figure}[!ht]
\includegraphics[width=\linewidth]{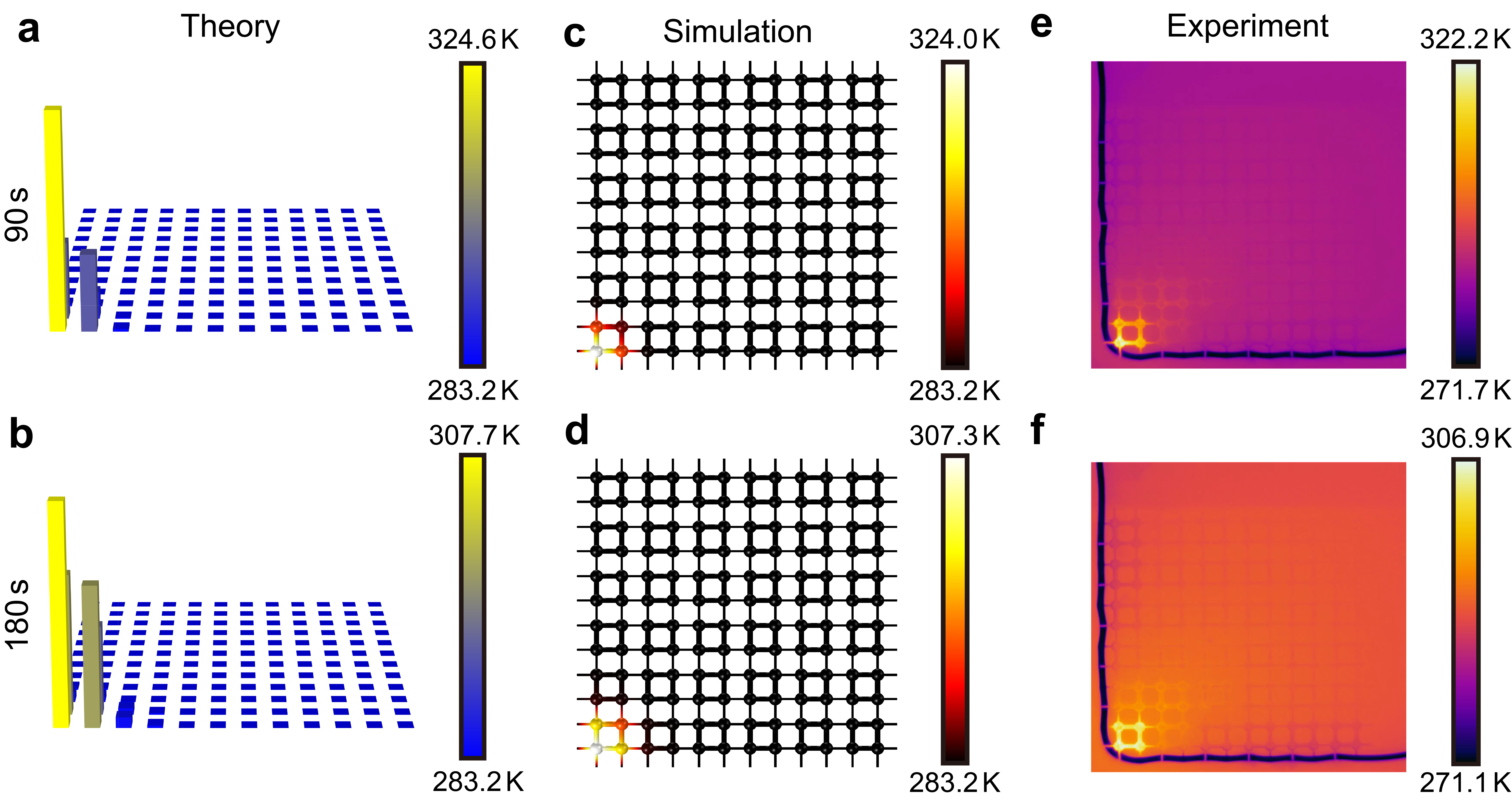}
\caption{Temperature distribution of trivial lattice. (a,b) Theoretical temperature distributions for the trivial lattice at (a) 90~s and (b) 180~s. (c,d) Simulated temperature distributions for the trivial lattice at (c) 90~s and (d) 180~s. (e,f) Experimental temperature distributions for the trivial lattice at (e) 90~s and (f) 180~s.}
\label{Fig3} 
\end{figure}

\clearpage
\newpage
\begin{figure}[!ht]
\includegraphics[width=\linewidth]{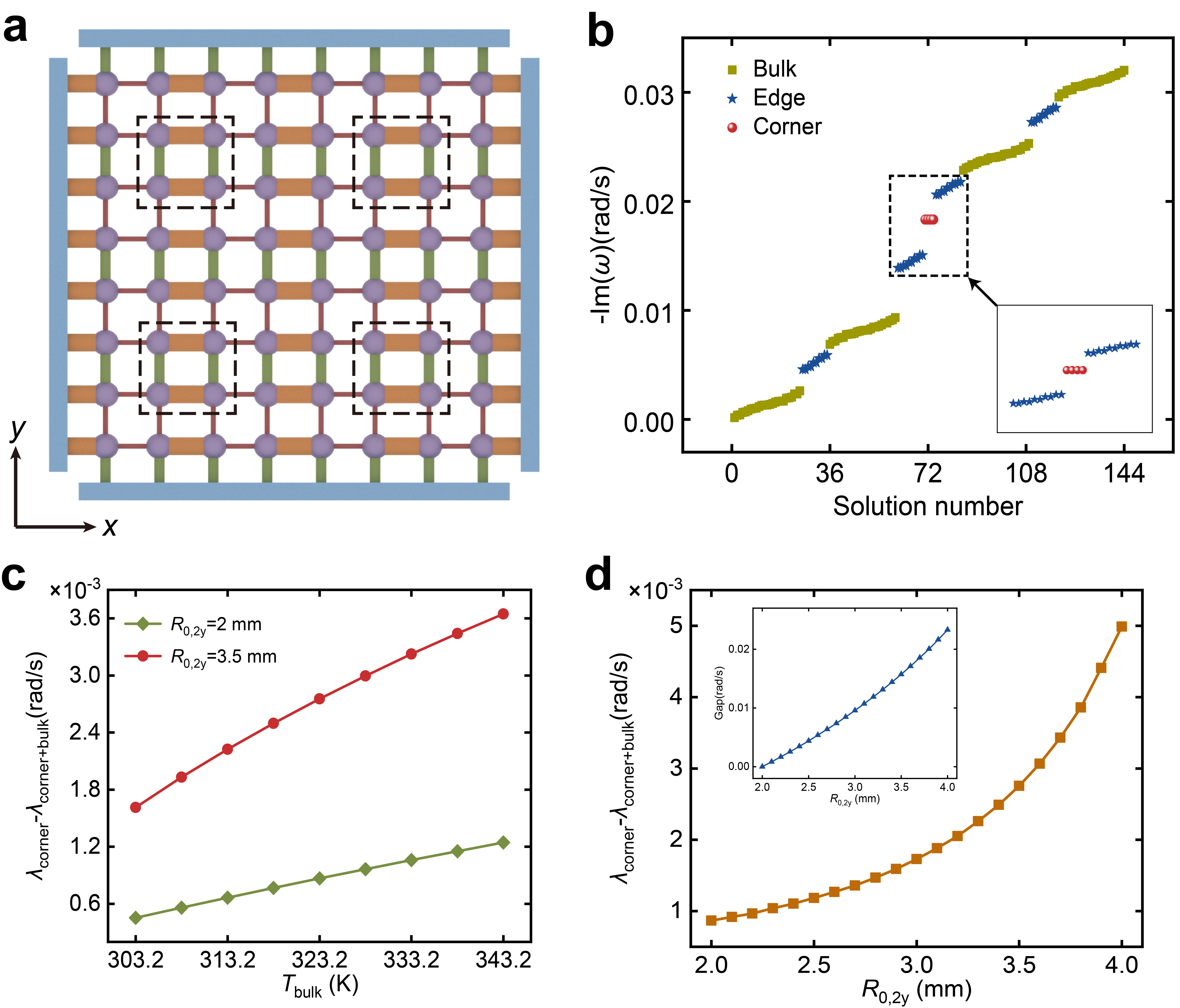}
\caption{In-bulk corner state and in-gap corner state. (a) Schematic diagram of sphere-rod structure for anisotropic diffusive 2D SSH model. Thermal diffusivities of red thin, green thick, and orange thicker rods are $D_{x1}=D_{y1}$, $D_{y2}$, and $D_{x2}$, respectively. The spheres in the black dashed boxes are stimulated as the bulk excitation. (b) The fixed boundary condition spectrum of anisotropic diffusive 2D SSH model. The inset is the enlarged view of four thermal corner states and their adjacent edge states. (c) The difference of decay rates (retrieved from the curve of normalized temperature with time) between corner spheres excitation and corner and bulk spheres excitation with different $T_{\rm{bulk}}$s. Here $T_{\rm{bulk}}$ is the excitation temperature of bulk spheres. $R_{0,2y}=3.5$~mm. (d) The difference of decay rates (retrieved from the curve of normalized temperature with time) between corner spheres excitation and corner and bulk spheres excitation with different $R_{0,2y}$s. The inset is the gap size between the second and third bands with the varying of $R_{0,2y}$. Here $T_{\rm{bulk}}=323.2$~K.}
\label{Fig4}
\end{figure}

\end{document}